\documentstyle[12pt]{article}
\begin{document}
\begin{titlepage}

\begin{flushright}
IASSNS-HEP-97/93\\
PUPT-1717\\
hep-th/9708037\\
August 1997
\end{flushright}

\vskip 0.2truecm

\begin{center}
{\Large {M(atrix) Theory on $T^6$ and a m(atrix) Theory Description
of KK Monopoles}}
\end{center}

\vskip 0.4cm

\begin{center}
{Amihay Hanany}
\vskip 0.2cm
{\it School of Natural Sciences\\
  Institute for advanced Study\\
Olden Lane, Princeton, NJ 08540 USA.\\
e-mail: hanany@ias.edu}

\vskip 0.2 cm
{and}
\vskip 0.2 cm

{Gilad Lifschytz}
\vskip 0.2cm
{\it Department of Physics,
     Joseph Henry Laboratories,\\
     Princeton University, \\
     Princeton, NJ 08544, USA.\\ 
    e-mail: gilad@puhep1.princeton.edu }

\end{center}

\vskip 0.8cm

\noindent {\bf Abstract} 
We discuss M(atrix) theory compactification on $T^6$. This theory
 is described by the large
$N$ limit of the world volume theory, of $N$ Kaluza-Klein monopoles 
in eleven dimensions.
We discuss the BPS states, and their arrangement in $E_{6}$ multiplets.
We then propose the formulation of the world volume theory
of KK monopoles in eleven dimensions that decouples from the bulk.
This is given by a 
large $N_1$ m(atrix) theory with eight supercharges, corresponding to the
quantum mechanics theory of $N_1$ zero-branes inside the 
Type IIA Kaluza-Klein monopole.
Various limits of the construction are considered.

\noindent

\end{titlepage}

\section{Introduction}

It has been realized in the last couple of years that all the superstring 
theories
in ten dimension are only some limits of a different theory, called M-theory 
\cite{wit1,ht}, 
that lives in eleven dimensions.
The superstring theories, as their
name suggests, were described by the interaction of strings, and  are
 the only known consistent quantum theories that include gravity.
When taking the strong coupling limit of the Type IIA string theory to obtain 
M-theory, the
strings became membranes. It might be that M-theory can be described
as a theory of interacting membranes, however such a description is not known.
Instead another description was proposed  \cite{bfss}, that M-theory in the
infinite momentum frame (IMF) is the large $N$, strong coupling limit of $SU(N)$
quantum mechanics, whose Lagrangian coincides with that of $N$ zero-brane at
short distances. This is known as M(atrix) theory. The M(atrix) theory 
description of M-theory
has by now passed many tests. It has the correct graviton graviton
scattering \cite{bfss,bb,bbpt} and the correct interaction between the extended
objects \cite{bss} of M-theory with \cite{pp}, and without
\cite{ab,gilmat,gil,ct,gilns,pier} eleven dimensional momentum
transfer. Taking the limit that
corresponds to weakly coupled Type IIA theory gives the correct description
of string interactions \cite{motl,dvv,bs}. 

M(atrix) theory compactification on $T^{d}$ have recently been extensively 
studied.
The low energy description of M(atrix) theory compactified on
 $T^{d}$ are given by a super Yang-Mills
theory in $(d+1)$ dimensions with sixteen supercharges, compactified on
 a dual torus $\tilde{T}^{d}$ \cite{bfss,tay}. For
$d \leq 3$ these theories are renormalizable and hence do
 not require extra information to be defined 
in the ultraviolet. For $d \geq 4$ the theories are not 
renormalizable and hence ill defined as 
they stand.

For all $d$ the behavior at strong coupling is of interest,
 as it is related in many beautiful ways to
known stringy phenomena \cite{setsus,sus,grt,rozali,fhrs}.  
For $d \geq 4$ this is also related 
to the ultraviolet behavior.

Clearly for $T^d$ with $d>3$ one has to add information how to treat
the ultraviolet regime. An obvious possibility is to regulate the
theories using string theory, by mapping the SYM theories to some brane
\cite{pol}
configurations in string theory. One however may suspect that this will 
necessarily introduce bulk effects (and in particular quantum gravity effects).
For example the theory
on the branes includes excitation of open strings stuck to the brane,
 in general two open strings can meet, become a closed string and leave the
brane. This process is the coupling of the bulk to the brane theory. If this
coupling is not suppressed then it seems unlikely we can describe the theory
on the brane by itself. This is only a problem for $d>3$.
 For $d=4,5$ the ultraviolet behavior has been
 suggested to be governed by 
the $(2,0)$ theory in six dimensions and by the theory of 
coincident NS-five branes at zero
string coupling respectively \cite{brs,seiberg}.
In these theories it was argued  that the bulk
decouples. In particular the theory of the NS-five branes at zero coupling, 
which
describes the compactification of the M(atrix) theory on $T^5$,  was
argued to be connected to the theory of non-critical strings living in 
$5+1$ dimensions \cite{dvv1,seiberg}, which do not include gravity.

In this paper we discuss
the compactification of M(atrix) theory on $T^6$. We propose 
that this is described by the theory of $N$ coinciding
Kaluza-Klein (KK) monopoles in eleven dimensions, in a limit
where the theory on the KK monopoles decouples from the bulk. The 
analog of the non-critical
strings of the NS-five brane theory are now membranes inside the world volume
of the KK monopole. Similar to the case of the string theories and M-theory, 
we
do not propose a description in terms of interacting membranes. Instead 
we propose a description (in the IMF) of that theory as a m(atrix) theory. 

In section (2) we review the compactification on $T^d$ with $d \leq 5$. In 
section
(3) we discuss the compactification on $T^6$ and the representation of the
U-duality group. In section (4) we propose a m(atrix) theory description of the 
theory of KK monopoles in eleven dimensions. Taking certain compactification
limits we get known descriptions of the theory supporting our proposal. We also
make a clear map between compactification of M(atrix) theory on $T^6$ and
the compactification of the proposed theory. We end with some speculations.

While this paper was finalized  \cite{egkr,lms,bk}  appeared, with some
overlap of ideas and results.


\section{Compactification: The General Picture}

The super-Yang-Mills prescription for compactification \cite{bfss,tay} breaks
down if the number of compactified direction is larger than three. 
This is because these theories are non-renormalizable, and thus ill defined.
One may hope to some how add some extra degrees of freedom to regulate
the theory in the UV, using string theory.
The idea is to match to the super Yang-Mills description at 
low energy a configuration of branes 
in string theory. From the knowledge of string dualities and
 behavior at large string coupling many
observations can be made on the strong coupling limit of those theories.
If these theories need extra information in the UV, than one has to take
various limits of parameters in order to decouple the bulk and still leave 
a non trivial theory on the brane. As every thing is embedded in string
theory the limit is a well defined theory (hopefully non trivial). 
Even 
before we worry about the decoupling from the bulk, the comparison to
brane configuration in string theory gives some information on these
theories.

First let us set-up some notation. Compactification of the M(atrix) theory
are defined by the parameters: the Planck length $l_{p}$, the radii of
the compact directions $L_{i}$, and $R$ which is the
 radius of the compact eleventh direction.
$R$ determines the coupling of the uncompactified ($0+1$ dimensional) theory,
\begin{eqnarray} 
g_{ym}^{2}  & = & g_{s} l_{s}^{-3}=\frac{R^3}{l_{p}^{6}} \\
g_{s} & = & R/l_{s}, \ \ \ \  l^{2}_{s}=\frac{l^{3}_{p}}{R}.
\end{eqnarray}
The M(atrix) theory compactification is  defined as the large $N$ $SU(N)$,
super-Yang-Mills (SYM) 
 theory, with sixteen supercharges, in $d+1$ dimensions on a dual torus
with radius ($\Sigma$), and coupling  ($g_{ym}$) given by (we drop factors of 
$2\pi$),
\begin{eqnarray}
\Sigma_{i} & = & \frac{l_{s}^{2}}{L_{i}}=\frac{l_{p}^{3}}{RL_{i}}. \\
g^{2}_{ym} & = & g_{s} l_{s}^{-3} \prod_i (\frac{l_{s}^{2}}{L_{i}})=
\frac{R^3}{l_{p}^{6}} \prod_{i}(\frac{l_{p}^{3}}{RL_{i}}).
\label{paraym}
\end{eqnarray}
As the super-Yang-Mills is just the theory of coincident
 branes in string theory 
these expressions can basically be derived from standard
 T-duality. The string coupling of the
system of coincident branes is then 
\begin{equation}
G_{s}=g_{s} \prod_i \frac{l_s}{L_i}.
\end{equation}

\subsection{Examples}

We now briefly review the known result of
 compactification up to $T^5$, stressing the qualitative understanding coming
from the relations to brane configurations in string theory.

\subsubsection{$T^2$}

We will start with M(atrix) theory on $T^2$. The Yang-Mills coupling is 
given by $g_{ym}^{2}=\frac{R}{L_{1}L_{2}}$. We model this by
 a theory of coincident D-2-branes
in a string theory with coupling $G_{s}$ and string length $l_{s}$ obeying
\begin{equation}
g_{ym}^{2}=G_{s} l_{s}^{-1}.
\label{gym2}
\end{equation}
We compare some energy scales
in the Yang-Mills theory and in the string theory. Take for
 example the non-threshold bound state
of a D2-brane and a D0-brane in string theory, and compare
 that in the appropriate approximation
to the energy of a magnetic flux in the Yang-Mills theory.
The energy of the bound state in the string theory is \cite{gilad, pol}
\begin{equation}
E_{s}=[(\frac{\Sigma_1 \Sigma_2}{G_{s} l_{s}^{3}})^2 +
 (G_{s} l_{s})^{-2}]^{1/2}
\label{e20}
\end{equation}
The energy in the YM theory is then the first term in the expansion
of equation (\ref{e20}) above the D2 brane mass.
One gets that the extra energy above the D2-brane energy is
 \begin{equation}
 \frac{l_{s}}{G_{s}\Sigma_1 \Sigma_2}.
\label{es}
\end{equation}
The energy of one flux of magnetic field in the YM theory is
\begin{equation}
E_{ym}=(g_{ym}^{2} \Sigma_1 \Sigma_2)^{-1}.
\label{eym}
\end{equation}
Comparing equation (\ref{es}) to equation (\ref{eym}) gives equation
(\ref{gym2}).

We now look at what happens when the YM coupling becomes large.
We think of the limit as a limit in M-theory.
This limit corresponds to fixing the eleven dimensional Planck length and
taking the string coupling to be large, thus the string length is also very
small.
This is the same as large $G_{s}$. A collection of D2-branes
at weak string coupling lives in ten 
dimensions that is why the YM theory has seven scalars representing
 the transverse excitations. But at 
strong string coupling the eleventh dimension opens up and
 the two-branes are free to move 
in a fully rotational invariant way in eleven dimensions 
so one expects at strong coupling which
in this case is the infra-red a fixed point with $SO(8)$
 symmetry \cite{setsus,bs}. The length of the new eleventh
dimension is
 \begin{equation}
R_{11}=G_{s} l_{s}=g_{ym}^{2} l^{2}_{s}=\frac{l_{p}^{3}}{L_1 L_2}.
\end{equation}

\subsubsection{$T^4$}
One models the M(atrix) theory on $T^4$ with a theory of D4-branes
in string theory with string coupling
$G_{s}$. This was analyzed in \cite{rozali}.
Similar consideration as before gives $g_{ym}^{2}=G_{s} l_{s}$.
Now we are interested
in the ultraviolet behavior which is like looking at the strong
 coupling behavior (as the YM coupling
has dimensions of length). In the Type IIA string as the string 
coupling becomes strong the eleventh 
dimension opens up. Unlike the case of the D2-brane, 
the D4-brane in string theory is wrapped around
the eleventh direction and is an M5-brane. The radius of 
the circle is just 
\begin{equation}
R_{11}=G_{s} l_{s}=g_{ym}^{2}=\frac{l_{p}^{6}}{RL_1 L_2 L_3 L_4 }
\label{4r11}
\end{equation}
Thus the ultraviolet theory is the theory of coincident M5-branes wrapped on
$T^5$ \cite{brs} with radii
$\Sigma_i$ given by equation (\ref{paraym}), and $R_{11}$
given by equation (\ref{4r11}).

\subsubsection{$T^5$}
The theory here is the theory of D5-branes.
The YM coupling and the string coupling are
\begin{eqnarray}
g_{ym}^{2} & = & \frac{l_{p}^{9}}{R^2 L_1L_2L_3L_4L_5}. \\
G_{s} & = & \frac{l_{p}^{6}}{R L_1L_2L_3L_4L_5}. 
\end{eqnarray}
The ultraviolet behavior is the 
strong coupling behavior. Now $g_{ym}^{2}=G_{s} l_{s}^{2}$,
 using S-duality of the string theory
this is a theory of NS-five branes in string theory with 
\begin{equation}
g_{NS}^{2}=G_{s}^{-1} (G_{s}^{1/2} l_{s})^2=l_{s}^2.
\end{equation}

Thus in the ultraviolet the string coupling goes to zero but 
leaving an interacting theory on the NS-five branes \cite{seiberg}.
It was argued in \cite{brs,seiberg}
that this theory contains non-critical string with finite tension, and
thus M(atrix) theory on $T^5$, may be described by a non-critical string theory.


\section{Matrix theory on $T^6$}

We model the theory by coinciding D6-branes in a string theory
 with coupling $G_{s}$.
The YM coupling is
\begin{equation}
g_{ym}^{2}=\frac{l_{p}^{12}}{L_1 L_2 L_3 L_4 L_5 L_6 R^3}
\end{equation}
But now one has the relation
\begin{equation}
g_{ym}^{2}=G_{s} l_{s}^{3}=L_{p}^{3}
\label{membrane}
\end{equation}
Where $L_{p}$ is the eleventh dimensional Planck length defined
 in the new string theory as 
\begin{equation}
L_{p}=G_{s}^{1/3} l_{s}.
\end{equation}

Now let us explore the ultraviolet region. Again this is just
the strong coupling region and the eleventh dimension opens up to size 
\begin{equation}
R_{11}=G_{s} l_{s}=\frac{l_{p}^{9}}{L_1 L_2 L_3 L_4 L_5 L_6 R^2}
\label{6r11}
\end{equation}
The other sides are
\begin{equation}
\Sigma_{i}= \frac{l_{p}^{3}}{RL_i}
\end{equation}
The M-theory interpretation of the D6-brane is as of a KK monopole
of eleven dimension \cite{town},
so M(atrix) theory on $T^6$ is described by the large $N$ limit of
$N$ KK monopoles in eleven dimensions.

Let us give a short description of some of the properties of
the KK monopole\footnote{For some recent discussion see \cite{hull,sen,bjo}.}. 
The multi- KK monopole is a gravitational background, in $n+4$ dimensions,
 of the form 
$R^n \times N_{4}$ where $N_{4}$ is the Euclidean 
multi-Taub-NUT metric \cite{haw}. It is given by,
\begin{eqnarray}
ds^2 & = &  -dt^2 +\sum_{i=1}^{n-1} d x_{i}^{2} + V^{-1}
(dx_{7} + \vec{w} d \vec{r})^{2} + V d \vec{r}^{2}.\\
V & = & 1 + \sum_{I=1}^{N} \frac{4A}{|\vec{r} - \vec{r}_{I}|}\\
\vec{\nabla}  \times \vec{w} & = & \nabla V. 
\end{eqnarray}
For this metric to be non singular $x_{7}$ must have periodicity of
$16\pi A$.  $\vec{r}_{I}$ denotes the location of the centers
of the different KK monopoles. We will refer to the compact direction $x_{7}$
as the NUT direction. 

The world volume theory of the KK monopole in eleven dimensions
is described by a vector multiplet with three scalars \cite{hull,sen1}. 
When two KK monopoles coincide there is an enhanced gauged symmetry coming
from extra massless states. The configuration of KK monopoles includes
a sphere of area proportional to the distance between the centers of the
KK monopoles. This is because the physical radius of the NUT direction goes
to zero at $\vec{r}=\vec{r_{I}}$. 
A two brane wrapped around the sphere has zero mass
in the limit in which the KK monopoles coincide, giving 
the extra massless states.

Let us now recover the compactification on $T^5$  by taking one of the sizes,
say $L_{6} \rightarrow \infty$. 
In this case $\Sigma_6 \rightarrow 0$ . Let us now take this 
direction as the coupling direction, the theory is well approximated by the 
theory of KK monopoles
in ten dimensional Type IIA string theory with coupling proportional to
$\Sigma_6$. The new string coupling becomes 
\begin{equation}
G_{s}^{new}=(\Sigma_{6} L_{p}^{-1})^{3/2}=
(\frac{L_1 L_2 L_3 L_4 L_5}{L_{6}^{2} l_{p}^{3}})^{1/2}.
\end{equation}
and the new string length is 
\begin{equation}
(l_{s}^{new})^{2}=(G_{s}^{new})^{-2/3} L^{2}_{p}=\frac{l_{p}^{9}}
{L_1 L_2 L_3 L_4 L_5 R^2}
\end{equation}
The KK monopole has five dimensional compact world volume directions with sizes
$\Sigma_{i}$ $i=1 \cdots 5$, and the Taub-NUT
direction is just $R_{11}$ from equation (\ref{6r11}).
This direction however is also very small 
in the limit $L_{6} \rightarrow \infty$, so we T-dualize
along this direction. We end up with a theory
on NS-five brane in type IIB, with string length $l_{s}^{new}$,
 the NS-five branes are
wrapped on the five-torus of sides $\Sigma_i$, there is an
 extra sixth direction which is compact 
given by T-dualizing $R_{11}$
\begin{equation}
L_{6}^{new}=\frac{(l_{s}^{new})^2}{R_{11}}=L_{6},
\end{equation}
and there is a new string coupling (because of the T-duality)
given by
\begin{equation}
G_{s}^{new}(IIB)=G_{s}^{new}\frac{l_{s}^{new}}{R_{11}}=
\frac{L_1 L_2 L_3 L_4 L_5  R}{l_{p}^{6}}
\end{equation}
The YM coupling on the NS-five branes is 
\begin{equation}
g^{2}_{NS}=G_{s}^{new}(IIB) (L_{s}^{new})^2 = \frac{l_{p}^{3}}{R}=l_{s}^{2}.
\end{equation}
Taking $L_{6} \rightarrow \infty$ 
we get  exactly the result in the $T^5$ case. Of course one does not have
to take that limit, and then  one gets the theory of NS-five branes with 
one transverse compact direction.

\subsection{BPS states and U-duality}

In this section we describe the BPS states and show they fall into the 
U-duality group representations. An  analysis of the moduli
space of M(atrix) theory compactification was done in \cite{bcd}. 

For M-theory on $T^6$ the U-duality group is $E_{6}$. 
What are the BPS states? Let us look at configurations of branes that can wrap
around some dimensions occupied by the KK monopole in eleven dimensions
and preserve some  of the original supersymmetries of M-theory. Let us
for convenience choose to work in Type IIA language, with
 The KK monopole becoming the  D6-brane. There are two different 
possibilities. First there are the bound states of the D6-brane and another
object, which preserves half of the original supersymmetries, they form
non-threshold bound states (from the point of view of the IIA theory), with mass
\begin{equation}
M^2=M_{6}^{2}+M_{D}^{2},
\end{equation}
where $M_{D}$ is the mass of the lighter object (we drop factors of $2\pi$) and
\begin{equation}
M_{6}=\frac{\Sigma_1 \cdots \Sigma_6 }{G_{s} l_{s}^{7}}=
\frac{V_6 }{G_{s} l_{s}^{7}}.
\end{equation}
In the YM theory they will correspond to states with energy
\begin{equation}
E=\frac{M_{p}^{2}}{2M_{6}}.
\end{equation}
The configurations are
\begin{enumerate}
\item 
An elementary string parallel to the D6-brane
forms a non-threshold bound state with it, there are $6$ such states.
For the elementary string
$M_{string}=\frac{\Sigma_{1}}{l_{s}^{2}}$, this gives for the Yang-Mills (YM) 
energy
\begin{equation}
E=\frac{\Sigma_{1}^{2} G_{s} l_{s}^{3}}{2V_{6}}=
\frac{\Sigma_{1}^{2}g_{ym}^{2}}{2V_{6}}.
\end{equation}
These can be represented in the YM theory as electric fluxes.

\item  A four brane forms a non threshold bound state, there are $15$ of those.
Now $M_{4}=\frac{\Sigma_{1} \cdots \Sigma_{4}}{G_{s} l_{s}^{5}}$, this gives
a YM energy of
\begin{equation}
E=\frac{(\Sigma_{1} \cdots \Sigma_{4})^2}{2V_6 G_{s} l_{s}^{3}}=
\frac{V_6}{2g_{ym}^{2}(\Sigma_5 \Sigma_6)^2 }. 
\end{equation}

These can be represented in the YM theory as magnetic fluxes.

\item A KK monopole of IIA string wrapped around the six dimensions. There
are $6$ such states.
If the Taub-Nut direction is $\Sigma_6$ then 
$$M_{kk}=\frac{\Sigma_1 \cdots \Sigma_5 \Sigma_{6}^{2}}{G_{s} l_{s}^{8}}.$$
This gives a YM energy of
\begin{equation}
E=\frac{(\Sigma_1 \cdots \Sigma_5 \Sigma_{6}^{2})^2}{2G_{s}^{3} l_{s}^{9} V_6}
=\frac{V_6 \Sigma_{6}^{2}}{2g_{ym}^{6}}.
\end{equation}

These are not easily represented in the YM theory. This is the same problem as
representing the transverse five-brane in M(atrix) theory (see also below). 

\end{enumerate}
These states transform under $SL(6,Z)$ as the $6$, $15$, and $6$ respectively.
Together these states gives  the $27$ of $E_{6}$.
Let us see how they transform into one another. Start with the state
of the D6-brane with world volume in directions $(1,2,3,4,5,6)$, bounded
to a D4-brane with world volume directions $(1,2,3,4)$. If we perform 
T-duality on directions $(2,3,4)$, then S-duality and then T-duality
again on directions $(2,3,4)$ we end up with the bound states of a D6-brane
and a fundamental string stretched in the $1$ direction. Similar
 transformation maps the bound state of the D6-brane and the D4-brane
 to the bound state of the D6-brane and the KK monopole of ten dimensions. 
So to get the $E_{6}$ one needs a transformation of
the form $(T^3 S T^3)$,
to be present in the regulated form of the world sheet theory\footnote{This 
kind of symmetry was used in \cite{grt,gilns}
to describe the NS-five brane in M(atrix) theory, in a very similar set-up.}.
This is just the uplifting of the S-duality of the $3+1$ dimensional 
YM theory (notice we always turn the D6-brane to a D3-brane by T-duality, then
S-dual and then T-dual again. This always gives us back a D6-brane).
A possible  manifestation of this 
symmetry in an ``improved'' SYM was recently
considered in \cite{egkr}.  

Second there are the bound states at threshold that preserve one quarter of
the original supersymmetries. The mass of the bound states is then just
\begin{equation}
M=M_{6} + M_{D}
\end{equation}
So the YM energy is just $M_{D}$. The states are
\begin{enumerate}

\item
A D2-brane wrapped on  any two of the six dimensions is
 a threshold bound state
with the D6-brane, there are $15$ such states. Now,
$M_{2}=\frac{\Sigma_1 \Sigma_2}{G_{s} l_{s}^{3}}$, so the YM energy is
\begin{equation}
E=\frac{\Sigma_1 \Sigma_2}{G_{s} l_{s}^{3}}=
\frac{\Sigma_1 \Sigma_2}{g_{ym}^{2}}.
\end{equation}
These can be described by instanton configurations in the YM, as can be seen 
from the coupling of the D6-brane world volume to background RR charges 
\cite{doug}.
Notice also that the membrane living inside the D6 brane worldvolume has a
tension, using equation (\ref{membrane}),
\begin{equation}
T_2\sim g_{ym}^{-2}=L_p^{-3}.
\end{equation}

\item
The NS-brane gives another $6$ threshold bound states.
Now $M_{NS}=\frac{\Sigma_1 \cdots \Sigma_5}{G_{s}^{2} l_{s}^{6}}$, this
then gives a YM energy of
\begin{equation}
E=\frac{V_6}{g_{ym}^{4} \Sigma_6 }
\end{equation}
These are not easily represented as a background. This is the same problem 
as for the KK monopoles above.

\item  A  graviton moving in one of the
 six directions gives $6$ states which are threshold bound states with 
the D6-brane. The YM energy is then  $E=\frac{1}{\Sigma_1}$.
These are just the KK modes in six dimensions.
\end{enumerate}
These states form a $\bar{27}$ of $E_{6}$. Again they transform to 
each other by $SL(6,Z)$ and a $T^3 S T^3$ transformation. The brane of the 
$27$ and $\bar{27}$ are
mapped into each other by electric magnetic duality in ten dimensions
(Hodge duality).  


\section{m(atrix) Theory of KK Monopoles}

In the previous section we made the observation that
M(atrix) theory on $T^6$ is the large $N$ limit 
of coincident KK monopoles in eleven dimensions. The low energy
 theory of the KK monopoles in eleven
dimensions is known and is described by a vector multiplet in seven
 dimensions. However  this can not be the whole story as it is
 ill defined.
We are faced with the prospect of understanding
the theory of KK monopoles in eleven dimensions, and finding the limit
of parameters where the bulk decouples.

We start by noting  an interesting analogy. 
In the case of $T^5$ it has been argued that the theory
includes non-critical strings, and may actually be formulated as a non-critical
string theory \cite{seiberg}.
For some possible constructions see \cite{abkss,wit3}.
This is also formulated as the theory of coinciding NS-five
branes. The compactification of M(atrix) theory on $T^6$ can be thought as
the strong coupling limit of coinciding D6 branes. 
If we compactify a transverse dimension to $N$ coinciding NS-branes, with 
vanishing string coupling, a series of S-dualities and T-dualities brings us to
the situation
of strongly coupled D6 branes. The string like excitation of the NS-five
brane theory are formally mapped to membrane excitations. All this sounds
much like the relation between ten dimensional string theories and M-theory.
In that case it is not clear that M-theory can be formulated as the theory
of membranes. However one candidate for the description of M-theory 
(in the IMF) is the BFSS M(atrix) theory. As we will see the analogy
is quite complete.

We propose a definition of the theory of coincident
KK monopoles in eleven dimensions (in the IMF) as the large $N$ quantum
mechanics of zero-branes with some extra matter restricted to the Coulomb
branch. This theory will be called
m(atrix) theory as compared with M(atrix) theory.

Imagine in the original M(atrix) model of BFSS we wish to add some p-brane.
 From the point of view of the zero-branes this is some background, for
instance a four-brane is described by turning on a background with
$\varepsilon^{ijkl} X_i X_j X_k X_l \neq 0 $ \cite{grt,bss}.  Alternatively,
if the p-brane zero-brane system preserves a quarter of the supersymmetries, 
then one can model this by adding more matter content that exactly
 reproduces the same effect. For example
in the case of  a D4-brane this is done by adding hypermultiplets \cite{bd}.
 So we end up with a quantum 
mechanics theory, with eight supercharges, 
describing the effect of the p-brane on the zero-brane.
 If one restricts to the branch where
the zero-branes are inside the p-brane (in the D4-brane it is the Higgs branch),
 this describes the theory on the 
brane as a m(atrix) theory (the theory should be in a limit
where the Higgs branch and the Coulomb branch are disconnected). 
In the large $N$, and strong coupling limit, the eleventh dimension
will open up and the D4-brane will become M5-brane. This was one of the 
approaches taken, to describe the 
$(2,0)$ theory as a m(atrix) theory \cite{abkss}. A similar suggestion was made
in \cite{dvv2}.

We wish to describe the KK monopole of eleven dimensions.
 We should start with a brane in ten dimensions with
the following properties. First it should become the KK
 monopole of eleven dimensions at strong Type IIA string coupling.
Second a configuration with extra zero-branes leaves a quarter of the
supersymmetries unbroken.
These two criteria are satisfied by the KK monopole in
 ten dimensions ( the D6-brane does not satisfy
the second criterion).

We propose that there is a  theory that lives on the world volume of the
KK monopole in eleven dimensions, and decouples from the bulk. This theory
is the restriction to the 
Coulomb branch, of the large $N_1$ limit of gauge
quantum mechanics, with eight supercharges, and some matter. Its 
Lagrangian coincides with
the Lagrangian describing the interaction  between $N_1$ zero-branes
inside the KK monopole of ten dimensions. 

The motion of the zero branes inside the KK monopole corresponds to the Coulomb
branch of the QM theory while the motion outside the KK monopole corresponds to
the Higgs branch of the QM theory.
We wish to decouple the theory on the KK monopole from the ten-dimensional bulk.
For this we need to decouple the Coulomb branch from the Higgs branch.
Since the mass of the hypermultiplets on the Coulomb branch is proportional to
the gauge coupling, in the strong gauge coupling limit the Higgs branch
decouples from the Coulomb branch.

The KK monopole in ten-dimensions has one 
special direction. The metric is of the form
$R^6 \times N_4$ where $N_4$ stands for the four
dimensional Euclidean Taub-NUT metric. One of the four
directions in $N_4$ is a circle (the NUT direction).
It is convenient to perform T-duality along this circle. The resulting
configuration consists of a collection of
NS-five branes and D-strings in Type IIB theory.
We choose the orientation of the branes as follows
\begin{eqnarray}
{\rm NS-five brane} & & \{0,1,2,3,4,5\} \\
{\rm D1-brane} & & \{ 0,6 \}
\end{eqnarray}

The decoupling of the Coulomb branch corresponds to the strong
coupling limit, which is like flowing to the infra-red. Thus we only need
the corresponding zero modes.
The zero modes of this theory can be deduced using the same reasoning
as in \cite{hw,witten}.

If we want to describe $m$ KK monopoles in eleven dimensions
we need to take $m$ NS-five branes on the circle $x_6$.
There will be $N_1$ D1-branes wrapped around the $x_6$ direction.
These in turn can break into segments stretched between each pair of adjacent
NS five branes.
The theory we are interested in is the theory
on the D1 branes in the large $N_1$ limit.
This theory contains 8 supercharges and is 0+1 dimensional. The R-symmetry for
this theory is $SO(5)\times SO(3)$ (or rather its extension to include spinor
representations) as will be explained below.

The matter content consists of a gauge group $U(N_1)^m$ with $m$
hypermultiplets
transforming in the bi-fundamental representations of each of the adjacent
factors in the gauge group. The $k$-th multiplet is transforming under the
$(N_1,\bar N_1)$ representation (with its complex conjugate) of
$U(N_1)_k\times U(N_1)_{k+1}$. There is cyclic symmetry such that the $m$-th
$U(N_1)$ group is coupled with a bi-fundamental representation to the first 
$U(N_1)$ group. The freedom in adding an arbitrary constant phase to the
position of the branes on the circle leads to
the decoupling of the $U(1)$ above. Similarly by translation invariance along
the (1,2,3,4,5) directions we get only $m-1$ moduli for the $U(1)$ factors of
the gauge group. This sets the sum over all these moduli to be zero.
A possible non-zero parameter can be introduced for the sum of these moduli
to be non-zero along the lines discussed in section 4 of \cite{witten}.
For $m=1$ the gauge group is $U(N_1)$ with matter in the adjoint
representation. A mass term for the adjoint can be introduced as for the
$m\not=1$ case. We will shortly see that this mass term is essential to
reproduce the correct result.

The presence of these branes break the ten-dimensional Lorentz symmetry into
two groups $SO(1,5)\times SO(3)$, where the first group acts on (0,1,2,3,4,5)
and the second group acts on (7,8,9).
These groups are identified as R-symmetries of the corresponding $0+1$
dimensional theory.
There are 4 parameters for a given NS five brane which describe its position in
the transverse space (6,7,8,9). The (7,8,9) position of the NS brane, which
transform as a vector of the $SO(3)$ group, parametrizes the FI couplings of
the $U(1)$ fields in the gauge group. A (7,8,9) distance between each two
adjacent
NS five branes is proportional to the FI coupling of the U(1) field in between
these two branes. The $x_6$ positions of the NS branes give the $0+1$
dimensional gauge couplings. Denote the $x_6$ position of the NS five branes by
$t_i$, $i=1,\ldots,m$. Then the gauge coupling of the $i$-th gauge group is
$g^2_i={g_s\over l_s^2|t_i-t_{i+1}|}$.

We can give an alternative description in the zero brane -- KK monopoles system.
The (7,8,9) positions of the KK monopoles are the FI couplings 
of the $U(1)$ factors
in the gauge group. In addition the QM gauge couplings correspond to periods of
the two form along the cycles given by the multi Taub-NUT solutions.

This theory was used to describe zero-brane near orbifolds, and 
M(atrix) theory in the presence of ALE singularity
in \cite{dgmo,dg,diagom}. There the analysis was focused on the Higgs branch,
while here we discuss the Coulomb branch.
In the analysis of \cite{dgmo,dg} they looked at
 the quantum mechanics on an  ALE space, the metric
on the Higgs branch is then the ALE metric. This is 
different than the metric of the Taub-NUT (which we might expect to get 
if the quantum mechanics we are discussing, is the one describing the
effect of the KK monopole in ten dimensions on zero branes). Both
metric have the same form but the defining function is
 $V=1+\frac{A}{r}$ in the Taub-NUT case and
$V=\frac{1}{r}$ in the ALE case. If however $A \rightarrow \infty$ we can
drop the $1$.
This is equivalent from our original point of view to
taking a KK monopole with a large circumference of $x_6$ which after
T-duality reduces to the case where the D1 brane is on a very small circle 
i.e. we have a theory in $(0+1)$ dimensions. This is however the limit
that we  are looking at, for the decoupling of the Coulomb branch from the
Higgs branch \footnote{
The vanishing of that term (or rather the limit $A \rightarrow \infty$)
was connected in \cite{lms,bk}
to the decoupling from  the gravity fields in the bulk.}

\subsection {Dynamics}

The description of the moduli space of vacua is as follows.
The following discussion should be taken in the context of the Born-Oppenheimer
approximation which serves as a pictorial description of the physics.
The ends of the
D1 branes can move inside the NS branes. The position of the D1 is given
by 5 scalars in the (1,2,3,4,5) directions which parametrize the Coulomb branch.
A motion into the Higgs branch can be done by going over to the Origin of the
Coulomb branch. A set of $m$ D1 branes can reconnect along the circle, each
between two adjacent NS branes, and form a
closed D1 brane which is wrapped around the $x_6$ direction. At this point the
D1 brane can move far from the system of NS branes along the (7,8,9) positions.
This is a transition to the Higgs branch where the (7,8,9) distance
parametrizes the expectation value for the hypermultiplets.
The transition from the Coulomb to the Higgs branch describes the
interaction of the KK monopoles with the bulk.

Another transition possible from the QM point of view is to turn on a FI term
for one of the $U(1)$ fields. This corresponds to moving the corresponding
NS brane away from the other NS branes along the (7,8,9) directions.
For this to happen all the D1 segments from both sides of the NS brane should
reconnect in such a way that no segments are left attached to the NS brane.
In field theory
this means that the $U(N_1)\times U(N_1)$ theory connected to this NS brane is
Higgsed by giving expectation value to the bi-fundamental hypermultiplet
localized on this NS brane.
The value of the FI term sets the expectation value for this hypermultiplet.
The resulting gauge group is $U(N_1)$ which is the diagonal embedding of these
two gauge groups. It is coupled to the other bi-fundamentals which are located
at the two ends of the D1 branes on neighboring NS branes.
We are left with a system of $m-1$ NS branes coupled to segments of $N_1$ D1
branes with a matter content just as for $m-1$ KK monopoles and $N_1$ D0 branes.
For the KK monopole -- D0 system this mechanism is just a decoupling of one
KK monopole from the other $m-1$ KK monopoles by moving it along the (7,8,9) 
directions.

It should be emphasized that the two transitions described above correspond to
effects which happen in the bulk of space time and does not correspond to the
description of the theory on the KK monopoles. This is since these transitions
are to or in the Higgs branch.
A necessary condition for having a Coulomb branch is to set the FI parameters
to zero and thus to the same (7,8,9) positions for the NS branes.

When all gauge couplings of the QM theory are infinite and there are no FI terms
the QM theory has a $SU(m)$ global symmetry. This symmetry is visible in the
brane construction by setting the positions of the NS branes along the (6,7,8,9)
directions to coincide. When the gauge couplings are finite this symmetry is
broken explicitly to $U(1)^{m-1}$. For this reason such a symmetry is not
visible in the Lagrangian formulation of the theory since it describes the
theory around the weakly coupled region. Only the diagonal part is visible.
In the IR limit, the couplings are sent to infinity
and thus we expect this symmetry to be visible in the spectrum (unless all
states are singlets of this symmetry). From the point of view of the KK monopole
theory this symmetry is expected. It just corresponds to the enhanced gauge
symmetry on the world volume of the KK monopoles when $m$ of them coincide.
For the eleven dimensional KK monopoles 3 moduli are needed to be tuned per one
KK monopole (the FI parameters), while for the ten-dimensional theory an
additional scalar is needed to be tuned (the gauge coupling).

Let us look at the case $m=1$. The QM theory contains a $U(N_1)$ gauge
theory coupled to hypermultiplets in the adjoint representation. There is a mass
term for the adjoint hypermultiplet coming from non-trivial boundary conditions
on the D1 branes ending on the NS brane \cite{witten}.
When the mass of the adjoint is zero the number of supersymmetries is enhanced
from 8 to 16. The QM theory then becomes just the BFSS proposal.
At this point the QM gauge theory will not be able to identify if the eleven
dimensional theory contains a KK monopole or not.
Thus the theories for $m=0$ (BFSS) and $m=1$ may appear the same.
However, a mass term for the adjoint field can be introduced only in the
presence of the NS brane. Only then there can be non-trivial boundary conditions
on the D1 branes which will allow for the non-zero mass.
In the limit in which the mass term is very large the adjoint field decouples
and we are left with a supersymmetric YM theory with gauge group $U(N_1)$ and
8 supercharges. So our proposal now includes another parameter in the 
gauge theory. Clearly for our proposal to be true this parameter can not be
zero.

If this parameter is not infinite there are two problems. First, it gives
another parameter that does not seem to exist in the theroy we are after.
Second, in order to have total decouling from the bulk, the D1 brane segments
should never be able to close, even at one point of the moduli space.
If the mass parameter is zero, the segments can join at the origin of the 
Coulomb branch, and thus there is no real total decoupling. 
If the mass parameter is non zero, there is no such point. However as we are
dealing with a $1+1$ dimensional field theory the moduli space is only in the
sense of a Born-Oppenheimer approximation. To ensure decoupling we should
then take the mass parameter to infinity. For the case $m=1$ this leaves
us with a large $N_1$  $U(N_1)$ QM gauge theory with eight supercharges.
 
The large $N$ limit of $U(N)$ QM gauge theory with 16
supercharges can be thought of as descrbing a membrane in eleven dimensions in
the lightcone frame \cite{dwhn}. A large $N$ 
$U(N)$ QM gauge theory with 8 supercharges
appears to describe a membrane in seven dimensions in the light cone frame 
(for a general review see \cite{dw}).
This fits nicely with our result, as the theory in $6+1$ dimensions should 
include a membrane.

Going back to the $m\not=1$ case. The bi-fundamental hypermultiplets receive
their mass from moduli on the Coulomb branch. Due to the special properties on
the $x_6$ circle, there are only $m-1$ such mass moduli.
A restriction comes from the sum of these masses to be zero.
The above discussion for the case $m=1$ then suggests that an additional mass
parameter needs to be introduced using the special boundary conditions on the
D1 branes. This mass parameter should probably be taken to infinity.


\subsection{Some Limits of m(atrix) Theory}

Let us consider some limits of this construction. This will Illuminate some
relations to other results, and strengthen our argument.
If the original KK monopole has $d$ compact directions
then the theory will be that of a D$(d+1)$-brane in
the presence of NS-five branes. The D$(d+1)$ brane is oriented along the $x_6$ 
direction and along $d$ more directions that are parallel to the NS-five brane.
In the $x_6$ direction the D$(d+1)$ brane
can break into segments stretched in between two adjacent NS branes.
The corresponding $d+1$ dimensional theory which describes the system of
KK monopoles with $d$ compact directions has 8 supercharges and contains
the same matter content as described in the previous subsection.

\subsubsection{The Non-Critical String Limit}

The m(atrix) theory we have just described can be thought of as describing the 
theory that
lives on the D6 branes in the IMF at strong Type IIA coupling
(actually infinite as the NUT direction is infinitely large).
If one of the directions of the D6-brane is compact and vanishingly small this 
can be T-dualized to give the theory on D5 branes at infinite coupling.
By S-duality this gives the theory
on Type IIB NS-five branes at zero string coupling. This theory was argued to be
non 
trivial and to contain non-critical strings \cite{seiberg}. 
In our construction this means that one of the NS-five brane directions
is vanishingly small and we can T-dualize it to get a system of NS-five brane in
Type IIA and a D2 brane
along $x_6$ and lets say $x_1$. In the infinite gauge coupling limit this
becomes a $(1+1)$ dimensional theory\footnote{For a recent analysis of the
Coulomb branch of some 1+1 dimensional field theories see \cite{diaseib}.}.

Our proposal is that the Coulomb branch
of this large $N_1$ theory describes the theory of non critical strings in
$(5+1)$ dimensions, in the IMF. This is the analogous construction to that of
Type IIA strings of \cite{dvv}.
In this case these strings are closed so they are mapped
to closed non-critical strings. 
Let us look at the case $m=1$. This will be a $U(N_1)$ gauge theory in
$1+1$ dimensions. Classicaly the moduli space is $(R^{4})^{N_1}/S_{N_1}$.
If this would be exactly correct one will reproduce the spectrum
of the Green-Schwartz superstring in six dimensions, which is not 
Lorentz invariant \cite{wit3}. However unlike the case of \cite{dvv} here 
we have only eight supercharges, one expects the moduli space to be 
corrected.
A possible description for Type IIA NS-five branes
at zero coupling was given in \cite{wit3} and also discussed in \cite{abkss}.

There is a possible mirror symmetry argument\footnote{A. H. would like to thank
discussions on this issue with Jacques Distler.}, to relate the present
construction to the construction of \cite{wit3,abkss}.
The theory on the Higgs branch which was studied in \cite{wit3} has a brane
description in terms of D2 branes stretched between D4 branes on the circle.
In the strong Type IIA coupling this system turns into membranes stretched
between five branes in M-theory. On the other hand the system of D2 branes
stretched between NS branes in Type IIA discussed above turns into the same
system in the strong Type IIA coupling. The M-theory system can not distinguish
from which system it has descended. This translates into mirror symmetry between
the two theories where the Higgs branch of one theory is mapped to the Coulomb
branch of the other theory and vise versa. Clearly many details need to be
worked out.

\subsubsection{The Weakly Coupled  D6 Limit}

Let us take the NUT direction to be very small.
The KK monopole in eleven dimensions can then be thought of as a D6 brane
in weakly coupled Type IIA 
boosted to infinite momentum along one of its world volume
directions. This should be described by open superstring
theory with some Dirichlet directions.

Now the set up is a collection of D1 branes wrapped around
a large $x_6$ direction and NS-five branes which
span directions $(1,2,3,4,5)$. 
In general the D1 branes will break in between the NS-five branes. 
The condition
that we restrict to the Coulomb branch is the condition that the D1 brane
can not move away from the NS-five branes. This can happen if some 
of the D1 brane segments join to form a circle,
and the D1 brane can then leave. 

The analysis of the boundary condition on the D1 brane segments
is the same as in \cite{hw}.  On the Coulomb branch 
the D1 brane can move inside the NS-five brane, but can  not
leave the NS-five brane, thus the boundary condition for the Coulomb branch
are Dirichlet boundary conditions in 
directions $(7,8,9)$ and Neumann in directions $(1,2,3,4,5)$. This is just
what one expects for an open string on a D6 brane in the IMF. Each D1 brane
segment has two labels which denote on which NS-five brane it ends.
They transform under the global hidden symmetries described earlier.
They correspond to the Chan-Paton factors
of the open string. One should construct the ``long strings'' in the
large $N$ limit, similar to \cite{motl,bs,dvv}. If the D1 branes join
and leave the NS-five brane (the Higgs branch), far along the flat direction
they will be approximately
described by just the world sheet theory of the D1 branes, which has the
right interactions to describe the closed Type IIA strings \cite{dvv}.


\subsubsection{Connection to the $(2,0)$ Theory}

If the KK monopole in eleven dimensions had two compact world volume directions
that are small and its NUT direction was also small than our theory
should reduce to the theory of Type IIA D4 branes in the IMF. In
our construction we see that we have a theory of $N$
D3 branes oriented along directions
$(1,2,6)$ and NS-five branes along directions $(1,2,3,4,5)$.
This is the theory that was originally studied in \cite{hw}.
The theory of Type IIA D4 branes in the IMF is then described by the Coulomb
branch
of the large $N$ limit of the theory on the D3 branes. However, this is related
by S-duality (mirror symmetry in the $2+1$ dimensional infra-red theory) to 
the Higgs branch of the large $N$ limit of D3 branes oriented
the same way together with D5 branes replacing the NS five branes.
The Higgs branch of this theory is the same as $N$ D1 branes oriented
orthogonal to D3 branes\footnote{This theory was analyzed in \cite{diac}.}. 

Now take the proposed theory to describe the M5 brane in the IMF \cite{abkss}.
It consists of the theory of large $N$ D0 branes and D4 branes. Suppose one of
the directions of the M5 brane, say $x^1$, which is not the IMF direction,
is compact and small. We can take this direction to be a string coupling
direction. This  reduces to the description of weakly coupled Type IIA
D4 branes in the IMF.
We can T-dualize along this compact $x^1$ direction to get a system of which
consists of $N$ D1 branes orthogonal to D3 branes.
This is the system which was discussed in the last paragraph.
This correspondence needs a further investigation. It suggests an interesting
connection, along the lines of mirror symmetry, between the two approaches for
defining the $(0,2)$ theory using the
Higgs branch and the KK monopole theory using the Coulomb branch.


\subsection{Back to M(atrix) theory on $T^6$}

When compactifying the BFSS M(atrix) theory on a torus we are led to theories 
that live on branes.
As was stressed in the introduction, one is looking for a theory that lives on 
the 
brane, is well defined and decouples from the bulk. We have given in 
the previous section a proposal for a m(atrix) description of the theory that 
lives
on the KK monopole of eleven dimensions, and decouples from the bulk, in
the infinite momentum frame. Now one needs to connect the variables of
the KK theory to those of M(atrix) theory on $T^6$. The parameters of the
theory living on the
KK monopole in the limit we have considered are the six lengths of the 
world volume, and the tension of the membrane that lives inside. The
analogous data of M-theory are the ten lengths and the tension
of the membrane which is $\sim l_{p}^{-3}$. In our m(atrix) model the length 
scale should then be $l_{g} \sim T_{membrane}^{-1/3}$.
Bellow this scale the theory is reasonably
described by $6+1$ dimensional SYM. Above this scale the SYM description breaks
down, and one has to use a different theory, the m(atrix) model. 
The mapping between the variables of the 
M(atrix) theory on $T^6$ and the theory of compactified KK 
monopole is as follows. The NUT direction is infinite.
This corresponds to the decoupling of the
Coulomb branch from the Higgs branch, and thus from the bulk.
The sides of the KK monopole are (as in section (3))
\begin{equation} 
\Sigma_{i}=\frac{l_{p}^{3}}{R L_i},
\end{equation}
and the tension of the membrane is (from section (3)):  
\begin{equation}
T_{membrane}^{-1}=L_{p}^{3}=g_{ym}^{2}=\frac{l_{p}^{12}}{L_1 L_2 L_3 L_4 L_5
L_6 R^3}=l_{g}^{3}
\end{equation}

This maps the variables of the M(atrix) compactification to the description 
given in section $(4)$.

There are problems with compactifying more than three directions in  our 
proposal. This
is because one ends up with a large $N_1$ gauge theory in more than four 
dimensions
which are not renormalizable. We hope our proposal is a first step in the right 
direction.


\section{Discussion}

In this paper we have analyzed the compactification of M(atrix) theory on $T^6$.
We were led to explore the theory living on the KK monopole of eleven 
dimensions.
We suggested a non perturbative description of the theory living on the
KK monopole that decouples from the bulk.
It is given by a m(atrix) description of
a large $N$, strong coupling limit, $SU(N)$ quantum mechanics with eight 
supercharges and appropriate matter content.  The analogy with the relationship
between string theory and the M(atrix) description of M-theory was stressed. 
As is the case with M(atrix) theory, the description of other branes 
(membrane, four-brane, etc), inside the seven dimensional world volume 
theory, can arise as some background. Their existence will be related to the
central charges of the supersymmetry algebra \cite{bss}.
It will also be interesting to do a
similar construction for KK monopoles of the Heterotic string. This 
presumably will give the analogs of the Heterotic M(atrix) model.

Compactifications of the theory were discussed, but it is not yet possible to 
get the
full description of M(atrix) theory on $T^6$.  This model however as it has a 
lot of similarity
to the BFSS model and its relation to the string theory may be useful in better
understanding
M(atrix) theory. Of-course this theory as a theory which presumably does not 
contain gravity is
of interest in its own right.

Let us end with some speculation. If one wants to describe certain theories in 
the IMF it seems possible to do that by taking the large $N$,
strong coupling limit of some supersymmetric $SU(N)$ quantum mechanics.
This was given in \cite{abkss} for the M5 brane and in this paper for the
KK monopole of eleven dimensions.
These results in a theory in $5+1$ and $6+1$ dimensions, respectively.
In order to end with a theory in $10+1$ dimensions (in the IMF)
one needs to start with a nine-brane
in Type IIA theory\footnote{Perhaps this is the nine-brane recently discussed
in \cite{hull}.}.
Then take the strong coupling limit to decouple from some un-known bulk,
and of course the large $N$ limit. Maybe this is a possible interpretation of 
the BFSS model.
The limits we need to take in the $(5+1)$ and $(6+1)$ dimensional 
theories, in order to decouple the bulk, are the exact same limits
we need to take in the BFSS model in order to get a unitary
Lorentz invariant theory. 

This may suggest (as has been suggested by others) that M-theory can 
be embedded in a higher dimensional theory.

\centerline{\bf{Acknowlegment}}
We would like to thank discussions with Ofer Aharony, Jacques Distler,
Savdeep Sethi, Eva Silverstein and Edward Witten.
The research of A. H. is supported in part by National Science Foundation grant
NSF-PHY-9513835.
The research of G. L. is supported in part by National Science Foundation grant
NSF-PHY-9157482.

\end{document}